\documentclass[aps,prb,showpacs,amsmath,amssymb,superscriptaddress,nofootinbib]{revtex4-1}
\usepackage{graphicx}
\usepackage{bm}
\usepackage{amssymb}
\usepackage{amsmath}
\usepackage{latexsym}
\usepackage{color}
\begin{document}  

\title{Material properties of $\alpha$-quartz that are relevant for its potential use in X-ray monochromators and analyzers}

\author{John P. Sutter}
\email{john.sutter@diamond.ac.uk}
\affiliation{Diamond Light Source Ltd, Harwell Science and Innovation Campus, Didcot, Oxfordshire OX11 0DE, United Kingdom}
\author{Hasan Yava\c{s}}
\affiliation{DESY, Notkestra{\ss}e 85, 22607 Hamburg, Germany}

\begin{abstract}
Monochromators and analyzers for X-rays of energy $>$ 5\ keV are overwhelmingly made from silicon crystals, of which large specimens of high purity and low defect density are readily available. Germanium and diamond crystals are also occasionally used. Bragg reflection of the X-rays determines their bandwidth, with backscattering reflections (with Bragg angles $\approx$ 90$^\circ$) providing the highest energy resolution along with high photon collection efficiency. The crystal lattices of all these materials share the same space group $Fd\overline{3}m$, which because of its high symmetry provides a low density of unique backscattering Bragg reflections per unit photon energy range. This is a disadvantage for solid-state studies using resonant inelastic X-ray scattering (RIXS) and nuclear resonant scattering (NRS), for which the X-ray energy must be tuned to a specific electronic or nuclear transition. As a result, crystals with lattices of lower symmetry are being considered. $\alpha$-quartz is one of the most promising. However, data on its material properties and fabrication techniques are widely scattered, and in the vital matter of the atomic positions, great inconsistencies exist in the literature. The material properties of $\alpha$-quartz that are relevant to X-ray studies are therefore summarized and clarified here as a guide to future work.
\end{abstract}

%
%

\maketitle

\section{\label{sec:INTRO} Introduction}

Precise studies of the atomic and nuclear properties of solid-state materials are often made with resonant inelastic X-ray scattering (RIXS)~\cite{RIXS1,RIXS2} and nuclear resonant scattering~\cite{NRS1,NRS2}. These techniques are applied to biophysics, chemistry, materials science, geophysics, and other fields. Energy bandwidths $<$ 10~meV for the incident X-ray photons and energy resolutions of similar magnitude in the analysis of the scattered photons are highly desirable. If the X-rays have energies over 5~keV, Bragg reflections from crystal monochromators and energy analyzers are by far the most suitable way to achieve these goals. The X-ray wavelength $\lambda$ selected by a set of diffracting atomic planes of a crystal is given by Bragg's Law, $\lambda = 2d\sin{}\theta_B$, where $d$ is the spacing between the atomic planes and $\theta_B$ is the grazing angle of incidence of the X-rays on the crystal (commonly known as the Bragg angle). (Note that the wavelength $\lambda$ is related to the photon energy $E$ by $E=12398.4~\textrm{eV$\cdot${\AA}}/\lambda$.) If $\Delta E$ is the bandwidth of the Bragg reflection and $\Delta\theta$ is the Bragg reflection's angular acceptance, the differentiation of Bragg's Law yields the relationship $\Delta E/E = \Delta\theta{}\cot{}\theta_B$, which indicates that the fractional energy bandwidth $\Delta E/E$ approaches zero as the Bragg angle approaches 90$^\circ$. In reality, detailed calculations using dynamical diffraction theory, which accounts for the coupling between incident and diffracted waves within a large perfect crystal, set a finite limit to $\Delta E$~\cite{DynDif}. Nevertheless, $\Delta E$ still drops to a level of several meV. At the same time, $\Delta\theta$ blows up from several $\mu$rad to as much as 1~mrad, thus improving the photon collection efficiency. However, the incident photon energy for a RIXS measurement must be close to the binding energy of a core-shell electron of a specific atomic species, and the incident photon energy for an NRS measurement must be close to a particular transition of the atomic nucleus being investigated. A crystal monochromator or analyzer can be used in backscattering at this photon energy only if it has a set of diffracting atomic planes with a spacing $d\approx 2\lambda$. The material of the crystal is therefore highly significant.

Large, pure and dislocation-free crystals of silicon are by far the most common material for X-ray monochromators and analyzers because they are widely available, thanks to the efforts of the semiconductor industry in refining the float-zone technique to manufacture meters-long ingots up to 150~mm in diameter. Germanium and diamond crystals are a distant second and third choice because of their lesser degree of lattice perfection, although germanium is sometimes chosen because its Bragg reflections have larger bandwidths than the corresponding reflections in silicon (thus passing more flux), and diamond has the advantage of very low X-ray absorption. All of these materials form crystal lattices with the same highly symmetric, face-centered cubic ``diamond structure'' space group $Fd\overline{3}m$. Unfortunately, the high symmetry greatly reduces the number of atomic plane sets with unique spacings $d$. Therefore, the number of unique backscattering Bragg reflections within a given range of photon energies is small, and --- unless one is very lucky --- the crystal monochromator and analyzer must generally be used quite far from backscattering in most RIXS and NRS measurements. Materials that form crystal lattices with lower symmetry, however, have a better chance of having a backscattering Bragg reflection close to a selected electronic or nuclear transition. Such alternative materials must be available in large ingots with low concentrations of impurities and defects, which distort the crystal lattice and hence broaden the energy bandwidth of the Bragg reflections. They must also be able to withstand cutting, polishing, and etching to the correct shape and size. They must be able to bend elastically without fracture, as many crystal analyzers for RIXS studies are spherically curved in order to capture more scattered photons. Finally, they must remain unaltered after long periods of X-ray bombardment. Synthetic, hydrothermally grown specimens of $\alpha$-quartz, the stable crystalline form of SiO$_2$ at room temperature and atmospheric pressure, have shown promising characteristics in the investigations carried out so far, which will be summarized below.

Despite the potential utility of $\alpha$-quartz, it proved difficult to collect and interpret all the most up-to-date data on its material properties because of the widely scattered sources. The positions of the Si and O atoms in $\alpha$-quartz, which must be known precisely in order to calculate $\Delta E$ and $\delta\theta$ for a given Bragg reflection, were especially problematic. Though many investigations of them exist in the literature, inconsistent choices of crystal lattice vectors have caused different authors to publish what at first appear to be very different coordinate values. Worse still, many authors do not explicitly define the axes that they used. This problem will therefore be handled first. The thermal properties of $\alpha$-quartz, which are important for precise setting of the X-ray photon wavelength and for use under high heat loads, will be treated next. A brief description of crystal fabrication and processing will then follow. Finally, a summary of tests and practical applications of $\alpha$-quartz crystal specimens will be provided.

\section{\label{sec:WgDpl} Material properties and crystal structure}

A detailed explanation of the properties of SiO$_2$ under various conditions is provided by Frondel~\cite{Silica}, and a useful brief review is provided by Johnson \& Foise~\cite{JoFoi}. The phase behavior of SiO$_2$ as a function of temperature and pressure is extremely complex; only a few pertinent facts are provided here. At atmospheric pressure, SiO$_2$ exists stably in the form of $\alpha$ or ``low'' quartz at temperatures up to 573$^\circ$C, of $\beta$ or ``high'' quartz from there up to 867$^\circ$C, of $\alpha$ tridymite from there up to 1470$^\circ$C, and of cristobalite from there up to the melting point of 1723$^\circ$C~\cite{Brice}. It should be stressed that various tridymite and cristobalite phases, as well as amorphous forms (glass) can exist metastably at room temperature and atmospheric pressure, but $\alpha$ quartz is the only thermodynamically stable form under normal ambient conditions. All of these forms have been found in nature. Other forms of SiO$_2$, such as keatite, coesite, and stishovite, have been synthesized under laboratory conditions and have since been found to exist under very rare conditions in nature.

X-ray and neutron crystallography has been applied to $\alpha$ quartz as to many other materials to determine the crystal structure and atomic positions. Crystals of $\alpha$ quartz have a trigonal Bravais lattice. The crystals are composed of SiO$_4$ tetrahedra that are linked together at their corners to form a three-dimensional network. These tetrahedra are slightly distorted so that one long type (1.6145~\AA) and one short type (1.6101~\AA) of Si-O bond exist~\cite{NtWl}. The conventional unit cell is hexagonal and contains three molecules of SiO$_2$. Its $c$-axis is a threefold screw axis; that is, a rotation of 120$^\circ$ about this axis followed by a translation of $+c$/3 along it leaves the lattice unchanged~\cite{Bragg}. If the 120$^\circ$ rotation appears clockwise when one looks along the negative $c$ direction, the screw axis is left-handed; if the rotation appears counterclockwise, the screw axis is right-handed. $\alpha$ quartz may exist in either of these forms, which are enantiomorphs (mirror images). $\alpha$ quartz crystals rotate the polarization of light propagating parallel to the $c$-axis, which is therefore also called the ``optical axis,'' in the same sense as the screw. In the first of many examples of the confusion plaguing the literature, $\alpha$ quartz with the right-handed screw is called \emph{levorotatory} (i.e. ``left'') quartz, and $\alpha$ quartz with the left-handed screw is called \emph{dextrorotatory} (i.e. ``right'') quartz~\cite{DoPa1}. There are also three twofold axes, all perpendicular to the $c$-axis. They are separated from one another by angles of 120$^\circ$ and intercept the $c$-axis at intervals of $c/3$. The absence of an inversion center allows $\alpha$ quartz to show piezoelectric effects when pressed along one of the twofold axes~\cite{BrGi}, which are therefore often named the ``electrical axes.''

Accurate knowledge of the atomic positions is indispensable for calculating X-ray structure factors, which determine the intensity and bandwidth of the Bragg reflections of $\alpha$ quartz. Many crystallographic studies have been published showing the atomic positions, but the inconsistencies between different papers make their interpretation difficult. Donnay and Le Page~\cite{DoPa1} discuss the following four areas of disagreement:

\subsection{Labeling of the space group}

The space group of the $\alpha$ quartz lattice is assigned the label $P3_121$ or $P3_221$ depending on the handedness of the $c$-axis. The International Tables for Crystallography define $P3_121$ as right handed and $P3_221$ as left-handed, and provide the special positions in a right-handed hexagonal coordinate system~\cite{InTbXl}. On the other hand, the diagram of the atomic positions in Wyckoff~\cite{Wyck} shows $P3_121$ defined as left-handed, though once again the hexagonal coordinate system is right-handed.

\subsection{Handedness of coordinate axes}

In older literature, the use of left-handed coordinate systems adds still more to the confusion. Lang~\cite{Lang} provides a worthwhile survey of the various conventions used by different authors. More recently, Donnay and Le Page~\cite{DoPa1} suggest that the handedness of the coordinate axes should be the same as that of the $c$-axis in order to allow a general treatment of both left-handed and right-handed $\alpha$ quartz.

\subsection{Rhombohedral setting: Miller indices of rhombohedral faces}

Quartz crystals have two rhombohedral faces, called ``major'' ($r$) and ``minor'' ($z$). If the $r$ and $z$ faces are assigned the Miller indices $(1~0~\overline{1}~1)$ and $(0~1~\overline{1}~1)$, respectively, the coordinates are said by Donnay and Le Page~\cite{DoPa1} to be in the $r$ setting. The alternate setting, labeled $z$, switches the Miller indices of the $r$ and $z$ faces.

\subsection{Piezoelectric orientation}

For a specified space group, coordinate system handedness, and rhombohedral setting, Donnay and Le Page~\cite{DoPa1} name the coordinates ``$(+)$'' if the twofold axis designated $a$ develops a positive charge at its positive end when stretched. If the charge on the positive end of the $a$-axis is negative, the coordinates are labeled ``$(-)$''. If the screw axis $c$ and the coordinate system have the same handedness, $r(+)$ and $z(-)$ are possible. If they have opposite handedness, $r(-)$ and $z(+)$ are possible.

In addition to all these, different authors disagree on the choice of the coordinate origin. They do consistently place it at the intersection of the $c$-axis and one of the twofold axes, and they also consistently orient the $a$ and $b$ axes of the hexagonal coordinate system along two of the twofold axes. However, Wyckoff~\cite{Wyck} gives the coordinates of the Si atom in the plane $c=0$ as $(u,0,0)$, whereas the International Tables of Crystallography~\cite{InTbXl} give them as $(-u,-u,0)$.

Bragg \& Gibbs~\cite{BrGi}, Wyckoff~\cite{Wyck}, and Kihara~\cite{Kihara} give consistent values for the lengths of the hexagonal axes $a$ and $c$. Kihara's values at 298~K are 4.9137~\AA and 5.4047~\AA, respectively. Baur~\cite{Baur} summarizes many previous measurements of these lattice parameters and provides his own, which are again close to Kihara's. In Wyckoff~\cite{Wyck}, the atomic positions of left-handed $\alpha$ quartz are given in the right-handed hexagonal coordinate system as follows:
\begin{itemize}
\item Si: $(u,0,0)$, $(\overline{u},\overline{u},\frac{1}{3})$, $(0,u,\frac{2}{3})$
\item O: $(x,y,z)$, $(y-x,\overline{x},z+\frac{1}{3})$, $(\overline{y},x-y,z+\frac{2}{3})$, $(x-y,\overline{y},\overline{z})$, $(y,x,\frac{2}{3}-z)$, $(\overline{x},y-x,\frac{1}{3}-z)$.
\end{itemize}
Wyckoff~\cite{Wyck} has $u=0.465$, $x=0.415$, $y=0.272$, and $z=0.120$. These are in good agreement with Le Page and Donnay~\cite{PaDo1} ($u=0.4699$, $x=0.4141$, $y=0.2681$, and $z=0.1188$) and the references therein, and also with Kihara~\cite{Kihara} at 298~K ($u=0.4697$, $x=0.4133$, $y=0.2672$, and $z=0.1188$). Kihara also shows $u$, $x$, $y$, and $z$ at seven higher temperatures up to 838~K, just below the transition to $\beta$ quartz, where $u=0.4855$, $x=0.4174$, $y=0.2397$, and $z=0.1422$. On the other hand, Baur~\cite{Baur} determines what at first seems like a completely different set of atomic positions for left-handed $\alpha$ quartz at 291~K: $u=0.5301$, $x=0.4139$, $y=0.1466$, and $z=0.1188$. Baur also cites a number of references that yield similar atomic positions. These results appear conflicting, but in reality the discrepancies arise only because different authors made different choices in the space group labeling, the handedness of the coordinate axes, the rhombohedral setting, the piezoelectric orientation, and the choice of origin. The atomic positions in each of the eight possible combinations of the first four of these factors are shown in Donnay and Le Page's~\cite{DoPa1} excellent table. Thus Wyckoff~\cite{Wyck} describes the case where the screw axis $c$ and the coordinate system have opposite handedness in the $z(+)$ setting. Baur~\cite{Baur}, on the other hand, uses the standardized form proposed by Parth{\'e} \& Gelato~\cite{PaGel}, according to which the Si atoms in right-handed $\alpha$ quartz are at positions $(\overline{u},\overline{u},0)$, $(u,0,\frac{1}{3})$, $(0,u,\frac{2}{3})$. After applying a 120$^\circ$ rotation about the $c$ axis, one sees that Baur has assumed that the screw axis $c$ and the coordinate system have the same handedness in the $z(-)$ setting.

\section{Thermal atomic motion, expansion, conductivity, and shock}

The temperature dependence of the structure factor of a Bragg reflection from a crystal is given by the reflection's Debye-Waller factor. To calculate this, one must know the mean square atomic displacements due to thermally induced vibrations. In the simple Debye model, these are dependent on the crystal's Debye temperature, which according to Gray~\cite{Gray} is 470~K for $\alpha$ quartz. However, experimental data indicate that the temperature dependence of the atomic vibrations is more complex than the Debye model. The thermal ellipsoids of the Si and O atoms in $\alpha$ quartz have been calculated by Young \& Post~\cite{YoPo}, by Lager, Jorgensen, and Rotella~\cite{LaJoRo}, and by Kihara~\cite{Kihara} as part of the least-squares refinement of X-ray and neutron diffraction patterns that produced their atomic positions and their lattice parameters. All measurements show that the thermal vibration is strongly anisotropic for both Si and O atoms. Although Le Page and Donnay~\cite{PaDo1} find that the mean square atomic displacements of the Si atoms are nearly isotropic, they also show considerable disagreement with previous work they cite. One should note that published values of the thermal ellipsoids are sensitive to extinction corrections and to the treatment of thermal diffuse scattering. Kihara~\cite{Kihara} describes in detail how the principal axes of the thermal ellipsoids rotate as temperature is changed, and shows the mean square atomic displacements along the principal axes as a function of temperature from 298~K to 838~K. According to this work, the thermal dependence of the mean square atomic displacements is not linear, increasing more rapidly at higher temperature.

The X-ray wavelength selected by crystal analyzers in backscattering is often adjusted by fine temperature control, which changes the spacing $d$ between the diffracting atomic planes. This is already common practice with silicon analyzers. The thermal expansion of $\alpha$ quartz is more complex that that of silicon because of its high anisotropy. Kosinski, Gualtieri \& Ballato~\cite{KoGuBa} analyzed the data published up to their time to produce a set of recommended linear thermal expansion coefficients from $-50^\circ$C to $+150^\circ$C. The $a$-axis coefficient increases monotonically with temperature over this range from $11.69\times 10^{-6}$~K$^{-1}$ to $15.55\times 10^{-6}$~K$^{-1}$. The $c$-axis coefficient increases monotonically with temperature over this range from $6.07\times 10^{-6}$~K$^{-1}$ to $9.04\times 10^{-6}$~K$^{-1}$. The recommended values at 20$^\circ$C are $13.55\times 10^{-6}$~K$^{-1}$ and $7.43\times 10^{-6}$~K$^{-1}$, respectively. Lager, Jorgensen \& Rotella~\cite{LaJoRo} provide values at 13~K, 78~K, and 296~K, as well as a useful survey of the thermal expansion coefficients at temperatures in between. The most remarkable feature is a sudden sharp drop in the $c$-axis coefficient, and a less pronounced drop in the $a$-axis coefficient, as $\alpha$ quartz is cooled below 80~K. Also remarkable is the negative $c$-axis thermal expansion below 12~K reported by White~\cite{White}.

Like thermal expansion, thermal conductivity in $\alpha$ quartz is highly anisotropic, being higher along the $c$-axis than the $a$-axis. Values recommended by the TPRC Data Series~\cite{TPRC} for thermal conductivity at 300~K are 0.104~W~cm$^{-1}$~K$^{-1}$ along the $c$-axis and 0.0621~W~cm$^{-1}$~K$^{-1}$ along the $a$-axis. At 500~K these are 0.060 ($c$) and 0.0388 ($a$) W~cm$^{-1}$~K$^{-1}$, and at 80~K (close to the boiling point of liquid nitrogen), these are 0.54 ($c$) and 0.279 ($a$) W~cm$^{-1}$~K$^{-1}$. The thermal conductivity reaches a maximum at 11~K along the $c$-axis and 10~K along the $a$-axis.

Any future high heat load applications should take into account the susceptibility of $\alpha$ quartz crystals to fracture caused by thermal shock. Brice~\cite{Brice} reports that $\alpha$ quartz crystals become more resistant as their infrared $Q$ increases (see below for definition). Typical thermal shock values sufficient to cause fracture are 40-80$^\circ$.

\section{Crystal fabrication and processing}

$\alpha$ quartz crystals appear in many modern appliances as resonators in electrical circuits. Brice's~\cite{Brice} review of the manufacturing, characterization, machining, and etching of these crystals may prove beneficial to the interested reader. Although $\alpha$ quartz crystals are common in nature, natural specimens generally have impurities, dislocations, and other defects that would degrade their performance as X-ray diffraction optics. Furthermore, they are subject to many different kinds of twinning, a phenomenon to which Frondel~\cite{Silica} devotes considerable attention. Finally, the supply of natural crystals has proven insufficient to fill the commercial demand. Synthetically grown $\alpha$ quartz crystals are more readily available, provide more consistently high crystal quality, and have consistent handedness (usually right). A fuller overview of the synthesis of $\alpha$ quartz is provided by Laudise~\cite{Laudi}. Unlike other materials, $\alpha$ quartz is not synthetically produced from melt, for several reasons. First, SiO$_2$ has a strong tendency to form glasses rather than crystals when cooled. Second, as previously mentioned, modified forms of high-temperature SiO$_2$ phases can continue to exist metastably at room temperature. Finally, SiO$_2$ crystals that pass from the $\beta$ quartz to the $\alpha$ quartz phase are prone to form twins. Therefore, the hydrothermal growth process is used instead. A ``nutrient'' of the desired material is dissolved at a high temperature. The warm solution is brought by convention currents into a cooler region and there becomes supersaturated. A ``seed'' crystal placed beforehand in the cooler region then takes up the excess solute molecules, incorporating them into its own crystal lattice. For $\alpha$ quartz, the nutrient is composed of natural quartz chips. A choice of two solvents is available: aqueous NaOH or aqueous Na$_2$CO$_3$ at 0.5-1~mol/L concentration. The nutrient, solvent, and seed are placed in a steel autoclave. The nutrient-containing lower region of the autoclave is heated to about 400$^\circ$C and the upper region, which holds the seed, to about 350$^\circ$C. These temperatures are slightly above and below the critical temperature of water, 374$^\circ$C. The pressure is well above the 220~bar critical pressure of water. Indeed, it may be as high as 2000~bars if NaOH is used, though it is less if Na$_2$CO$_3$ is used. Tokyo Denpa (recently acquired by Murata) has been a supplier to the synchrotron community; Ao \emph{et al}~\cite{Ao} also name Inrad and Ecopulse as suppliers.

Inclusions, produced either by trapped solvent or by specks of complex metal-silicate compounds formed on the inner wall of the autoclave, may appear in synthetically grown crystals, especially around the seed. Dislocations may originate either from inclusions or from the seed crystal itself. Therefore, careful selection of the seed crystal is vital, as etchants preferentially attack the strongly strained material around dislocations, and in this way can produce channels passing right through a plate. The main impurity is hydrogen (believed to be bound in hydroxyl ions). The concentration of hydrogen is tested by infrared absorption. Lithium and sodium impurities are kept low because of their damaging effects on the crystal's electrical properties. Aluminum impurities make $\alpha$ quartz crystals susceptible to damage by ionizing radiation. Manufactureres such as Tokyo Denpa typically provide the following specifications: absence of twinning, low inclusion density, low etch channel density, and good infrared absorption. Crystals 50~mm in diameter and 2~mm thick have been obtained by the Advanced Photon Source~\cite{Diego}. In Brice~\cite{Brice}, all these parameters, including the angular width of X-ray Bragg reflections, are shown to improve with increasing ``infrared $Q$,'' which is defined as the electrical quality factor of an $\alpha$ quartz resonator manufactured in a standardized way.

$\alpha$ quartz crystals are sometimes ``swept'' after fabrication by electrolysis at a temperature just below the transition to $\beta$ quartz. This removes interstitial ions in the crystal lattice or replaces them with protons, holes, or insoluble species. By this process, the radiation hardness of the crystal is increased and the likelihood of etch channels is reduced~\cite{JoFoi}.

$\alpha$ quartz crystals may be crystallographically oriented to a particular Bragg reflection using standard X-ray equipment. So too, cutting and lapping may be done using conventional diamond wheel saws and common abrasives, such as alumina or silicon carbide. Etchants include ammonium fluoride (NH$_4$F) and ammonium bifluoride (NH$_4$HF$_2$), hydrofluoric acid, and sodium hydroxide. Hedlund et al~\cite{Hedlun} have examined etching rates of various aqueous HF/NH$_4$F mixtures at temperatures from 22$^\circ$C to 80$^\circ$C. Strong anisotropy of the etch rate is reported. Like other materials, particularly silicon, $\alpha$ quartz crystal wafers can be diced and attached to a substrate (generally of glass) to produce a curved backscattering high-energy-resolution analyzer capable of X-ray focusing. Glue is often used to bind the crystal to the substrate. However, direct bonding methods, which rely not on adhesives but on the formation of chemical bonds between the crystal and the substrate, are being actively researched by developers of microelectical-mechanical systems (MEMS) and microfluidics. Quartz-to-quartz~\cite{Rang}, quartz-to-silicon~\cite{Eda}, and quartz-to-glass~\cite{Naka} techniques have been published. Direct bonding requires very flat and highly polished surfaces with a micro-roughness $<$ 5~\AA.

\section{Tests and practical applications with X-rays}

Measurements of the degree of imperfection in the atomic lattices of unbent $\alpha$ quartz crystals have been carried out numerous times. H{\"o}nnicke \emph{et al}~\cite{Hoenn} provide a list of references, and also measured relative variations of $\Delta d/d\approx 5\times 10^{-7}$ of the lattice spacing $d$ of the $(\overline{4}~2~2~0)$ Bragg reflection over areas of 79~mm$\times$32~mm. Investigations of X-ray backscattering from unbent $\alpha$ quartz crystals are reported by Sutter \emph{et al}~\cite{Sut1} on the $(11~6~\overline{17}~0)$ Bragg reflection at 21.747~keV and again~\cite{Sut2} on the $(7~\overline{4}~\overline{3}~4)$ Bragg reflection at 9.979~keV. Measured values of the energy width of the latter reflection were below 4~meV. Imai \emph{et al.} began their investigations beforehand, but their published report, a measurement of backscattering by the $(0~6~\overline{6}~10)$ Bragg reflection of 14.4~keV X-rays emitted by the $^{57}$Fe M{\"o}ssbauer resonance, appeared slightly later~\cite{Imai}. They measured a bandwidth of 1.14~meV. 

Detailed theoretical comparisons of spherical diced backscattering analyzers of silicon, germanium, $\alpha$ quartz, sapphire, and lithium niobate have been compiled by Gog \emph{et al}~\cite{Gog}. Lider, Baronova \& Stepanenko~\cite{Lider} observed a periodic deformation both of 150~$\mu$m thick silicon crystals glued to spherical aluminum substrates and of 80-100~$\mu$m thick $\alpha$ quartz crystals bonded without glue to spherical, toroidal, or cylindrical fused-quartz substrates of nominal radii 0.18-3.4~m. They attribute this to relaxation of elastic stresses and point out the significant broadening of energy resolution that would result. S{\'a}nchez del R{\'i}o \emph{et al}~\cite{Sanch}, on the other hand, found good agreement between their experimental measurements of spherically bent 60~$\mu$m thick $\alpha$ quartz crystals at the European Synchrotron Radiation Facility and theoretical calculations using a multilamellar model, but found it necessary to adjust the Debye-Waller factors and the crystal thickness in their model. Their radii of curvature were 150~mm and 250~mm. Ao \emph{et al}~\cite{Ao} compared the collection efficiency and spectral resolution of $\alpha$ quartz crystal analyzers spherically bent to a 150~mm radius with mica, germanium, and pyrolytic graphite analyzers. Quartz compares favorably to the other materials in resolution. Pereira \emph{et al}~\cite{Pereir} performed monochromatic X-ray topography at the Advanced Photon Source on $\alpha$ quartz crystals spherically bent to a 672~mm radius. Although the results were encouraging, they also observed localized defects that had no evident connection with visible imperfections in the crystal and did not appear confined to the crystal surface, but did differ depending on whether the diffracting surface of the crystal was $(1~1~\overline{2}~0)$ or $(1~0~\overline{1}~1)$. Finally, Ketenoglu \emph{et al}~\cite{Keten} manufactured a curved diced $\alpha$ quartz analyzer oriented to the $(2~4~\overline{6}~4)$ lattice plane and obtained a 25~meV energy resolution at 8981~eV.

\section{Conclusions}

This report is intended to introduce those interested in designing high-energy-resolution $\alpha$ quartz crystal X-ray monochromators and analyzers to the relevant literature. Though the available publications are too numerous for this brief paper to provide a complete list, an attempt has been made to bring the reader up to the current state of the art in the application of $\alpha$ quartz to measurements of electronic properties in solid-state materials using inelastic X-ray scattering. In addition, several references on the thermal conductivity and thermal shock resistance have been provided for future researchers who may use $\alpha$ quartz crystals under high radiation bombardment. The references contained in the cited papers will help interested readers to learn more details if they wish. The most recent quality tests and practical applications of $\alpha$ quartz backscattering analyzers have been provided at the end for completeness. The development of $\alpha$ quartz X-ray analyzers, and also of X-ray analyzers made from sapphire and lithium niobate (also trigonal crystals), is ongoing.

\section{Acknowledgments}
The authors thank Stanislav Stoupin (formerly of the Advanced Photon Source, now at Cornell University) and Nino Pereira of Ecopulse for stimulating discussions about $\alpha$ quartz. They also thank Diego Casa of the Advanced Photon Source for his advice. Alfred Baron deserves special mention for his insightful questions, which led directly to the writing of this report.

\end{document}